\documentstyle[12pt]{article}

\begin{document}
\baselineskip 22pt
\title{
Discussion: Time-Symmetric Quantum Counterfactuals}
\author{ Lev Vaidman}
\date{}
\maketitle

\begin{center}
{\small \em School of Physics and Astronomy \\
Raymond and Beverly Sackler Faculty of Exact Sciences \\
Tel Aviv University, Tel-Aviv 69978, Israel. \\}
\end{center}

\vspace{2cm}
\begin{abstract}\baselineskip 22pt
  There is a trend to consider counterfactuals as invariably
  time-asymmetric. Recently, this trend manifested itself in the
  controversy about validity of counterfactual application of a
  time-symmetric quantum probability rule. Kastner (2003) analyzed
  this controversy and concluded that there are time-symmetric
  quantum counterfactuals which are consistent, but they turn out to
  be trivial. I correct Kastner's misquotation of my defense of
  time-symmetric quantum counterfactuals and explain their non-trivial
  aspects, thus contesting the claim that counterfactuals have to be
  time-asymmetric.
\end{abstract}

%\vfill\break

\vskip 1.8cm
 \noindent
 {\bf 1. Introduction.~~ }
\vskip .6cm

 The issue of time (a)symmetry of
 counterfactuals has been addressed many times in the past, but it
 remains to be an open question, see for example Kutach (2002). Less
 than a decade ago, Sharp and Shanks (1993) opened the discussion of
 time-symmetric counterfactuals in the context of quantum theory
 (TSQC), claiming that time-symmetric approach of Aharonov, Bergmann,
 and Lebowitz (ABL) (1964) to quantum measurements is not applicable
 to counterfactual situations. The controversy which aroused after this
 paper was reviewed by Kastner (2003) who proposed that the ABL TSQC
 can be considered consistent, but that they provide no new
 information. In her analysis she presented my defense of the TSQC,
 but she misquoted and misinterpreted  it. Here I want to correct this,
 to state what is the difficulty of TSQC which I resolved,  and
 contest the claim that the TSQC are trivial.

%\footnote{I will not
% consider Kastner's criticism of the Mohrhoff (2000) approach.}

The plan of the paper is as follows. In the next two sections I describe the two misquotations of my
approach made by Kastner. In Section 4 I briefly discuss the
controversy about counterfactual application of the ABL rule and in
sections 5 and 6 I analyze two examples. Section 7 concludes the paper by discussion of the relation
to the general question of time-symmetry of counterfactuals. 

\vskip 1.8cm
 \noindent
{\bf 2. First Misquotation.}
\vskip .6cm

In her arguments, Kastner (2003) relies few times on the fact that  I
`acknowledge that post-selection results can't be actually ``fixed'''
(p.8,23 of preprint). To support this claim she brings in footnote 5 a
quotation from Vaidman (1999a). Let me enlarge the 
 quotation, the part
quoted by Kastner appears in the second paragraph and I put it in italics.

\begin{quotation}
 A different asymmetry
(although it looks very similar) is in what we assume to be ``fixed'',
i.e.,  which properties of the actual world we assume to be true in
possible counterfactual worlds.  The {\it past} and not the {\it
  future} of the system is fixed.

It seems that while the first asymmetry can be easily removed, the
second asymmetry is unavoidable. According to standard quantum theory
a system is described by its quantum state. In the actual world, in
which a certain measurement has been performed at time $t$ (or no
measurement has been performed at $t$) the system is described by a
certain state before $t$, and by some state after time $t$. {\it In the
counterfactual world in which a different measurement was performed at
time $t$, the state before $t$ is, of course, the same, but the state
after time $t$ is invariably different (if the observables measured in
actual and counterfactual worlds have different eigenstates).
Therefore, we cannot hold fixed the quantum state of the system in the
future.}\footnote {\baselineskip 20pt Note that none of these asymmetries exists in the
  classical case because when a complete description of a classical
  system is given at one time, it  fixes the complete
  description at all times and (ideal) measurements at time $t$ do not
  change the state of a classical system.}

The argument above shows that for constructing time-symmetric
counterfactuals we have to give up the description of a quantum system
by its quantum state. Fortunately we can do that without loosing
anything except the change due to the measurement at time $t$ which
caused the difficulty. A quantum state at a given time is completely
defined by the results of a complete set of measurements performed
prior to this time.  Therefore, we can take the set of all results
performed on a quantum system as a description of the world of the
system instead of describing the system by its quantum state. (This
proposal will also help to avoid ambiguity and some controversies
related to the description of a single quantum system by its quantum
state.) Thus, I propose the following definition of counterfactuals in
the framework of quantum theory:

\begin{quotation}
{ \bf (ii)}
{\em  If  a measurement ${\cal M}$  were performed at
  time $t$, then it would have property ${\cal P}$, provided  that the results of all measurements
performed on the system at all times except the time $t$ are fixed.}
\end{quotation}

For time-asymmetric situations in which only the results of
measurements performed before $t$ are given (and thus only these
results are fixed) this definition of counterfactuals is equivalent to
the counterfactuals as they have usually been used. However, when the
results of measurements performed on the system both before and after
the time $t$ are given, definition (ii) yields novel time-symmetrized
counterfactuals. In particular, for the ABL case, in which {\em
  complete} measurements are performed on the system at $t_1$ and
$t_2$, $t_1 < t <t_2$, we obtain
\begin{quotation}
  { \bf (iii)} {\em If a measurement of an observable $C$ were
    performed at time $t$, then the probability for $C=c_j$ would
    equal $p_j$, provided that the results of measurements performed
    on the system at times $t_1$ and $t_2$ are fixed.}
\end{quotation}

\end{quotation}

Just from the structure of my writing, it is clear that I do not claim
as true what  Kastner took as the quotation: the paragraph starts
with ``It seems'' and in the following paragraph I show that we can
overcome the difficulty. Moreover, as I will explain below, the difficulty is
not related to fixing the outcome of the measurement at $t_2$, the
issue which concerns Kastner.

\vskip 1.8cm
 \noindent
{\bf 3. Second Misquotation.}

\vskip .6cm Second misquotation is Kastner's claim that I ``attribute
values to observables that were not measured''. Indeed, the name
``elements of reality'' (Vaidman 1993) and the title ``How to
Ascertain the Values of $\sigma_x$, $\sigma_y$, and $\sigma_z$ of a
Spin-$1\over 2$ Particle'' (Vaidman, Aharonov, and Albert, 1987) might
suggest this.  However, if Kastner wants faithfully to present my
approach, she should not ignore my reply (Vaidman, 1999c) to her other
paper (Kastner, 1999b).  Let me quote two paragraphs from my reply (p.
866).

\begin{quotation}

I {\it define} that there is an element of reality at time $t$ for an
observable $C$, ``$C=c$'' when it can be inferred with certainty that
the result of a measurement of $C$, if performed, is $c$. Frequently,
in such a situation it is said that the observable $C$ has the value
$c$. It is important to stress that both expressions do not assume
``ontological'' meaning for $c$, the meaning according to which the
system has some (hidden) variable with the value $c$. I do not try to
restore realistic picture of classical theory: in quantum theory
observables do not possess values.  The only meaning of the
expressions: ``the element of reality $C=c$'' and ``$C$ has the value
$c$'' is the operational meaning: it is known with certainty that if
$C$ is measured at time $t$, then the result is $c$.

Clearly, my concept of elements of reality  has its
roots in ``elements of reality'' from the Einstein, Podolsky, and Rosen
paper (EPR) (1935). There are numerous works analyzing the EPR elements of
reality. My impression that EPR were looking for an ontological
concept and their ``criteria for elements of reality'' is just a
property of this concept. I had no intention to define such ontological
concept. I apologize for taking this name and using it in a very
different sense, thus, apparently, misleading many readers. I hope to
clarify my intentions here and I welcome suggestions for alternative
name for my concept which will avoid the confusion. 
\end{quotation}

If there is an element of reality $C=c$, then, apart from the
counterfactual statement about the result of the measurement of $C$
(which is the definition of ``element of reality''),
the quantum system has some other features, as  will be described in
two examples in Sections 5 and 6. However, it does not mean that there
is something in the system which possesses value $c$.

\vskip 1.8cm
 \noindent
{\bf 4. The controversy about counterfactual application of the ABL rule.}
\vskip .6cm

The ABL rule is usually considered in  situations in which the
counterfactual has compound antecedent with three parts: (1) result
of a complete measurement at $t_1$, (2) the fact that some measurement was
performed at time $t$, and (3) the result of a complete measurement
at $t_2$, $t_1<t<t_2$. I have argued before (1999a) that the controversy
aroused from the error of Sharp and Shank, who considered the
probability of the result of a measurement at time $t$ without taking
in account (2), i.e., the fact that the measurement has been
performed. 

Kastner admits that when we take all compounds of the antecedent, the
inconsistency proof of Sharp and Shanks fails, but claims that the
counterfactuals in this case are trivial, not surprising, and do not
yield any new information. Her argument is based on the analogy with
a classical example of a counterfactual with compound antecedent which
looks surprising when only one compound is fixed and which 
trivially holds when all compounds are taken into account. The
surprising property of the classical counterfactual is explained then by
small probability of having all compounds of the antecedent true
together.

It is hard to accept Kastner's  argument according to which if two
statements have the same form and one is trivial (Kastner's raffle
example), then the second (counterfactual application of the ABL rule)
must be trivial too.  Let me spell out in the next two sections what are, in my view,
nontrivial surprising features of the ABL counterfactuals and what is the new information which
we can get from them analyzing examples mentioned in
Kastner's paper.

%\break
\vskip 1.8cm
 \noindent
{\bf 5. Elements of Reality for a spin-$1\over 2$ particle.}

\vskip .6cm

I will convert the first example into a raffle. In a raffle each
participant brings his own spin-$1\over 2$ particle on which the
organizers perform a measurement of a spin component in one of the
there directions, $x, y,$ or $z$. The participant gets his particle
back and he has to provide three statements: if the measurement was in
$x$ direction, the result was $s_x$, if the measurement was in $y$
direction, the result was $s_y$, and if the measurement was in $z$
direction, the result was $s_z$. (This is a typical situation when
counterfactuals considered in the context of quantum mechanics.
Several statements are made together about measurements which cannot
be performed together.) The participant is a winner, if his statement
about the measurement which was actually performed was correct.

Our surprising result is that a participant equipped with quantum
devices can always win. It is simple to prepare the spin in one of the
three directions. The participant can also measure the spin in another
direction when he gets the particle back. In this way he can infer the
results of measurements in two directions, but to know the results
for three directions is a highly nontrivial task. Experimentalists
found it interesting enough to actually perform this experiment in a
laboratory, Schulz (2002).

In this example there is no small probability of having all
compounds of the antecedent true together. To achieve the task, the participant have
to perform certain measurements which might have different outcomes,
but in {\em all} cases he can make correct statements.  (It does not
mean that I accept Kastner's argument about ``cotenability'' of the
pre- and post-selection together with a particular intermediate
measurement. The fact that in many ``surprising'' situations the
probability to succeed in the post-selection is small, does not make
corresponding counterfactuals vacuous as Kastner claims. The only
requirement is that the probability for the post-selection does not
vanish.)

One might claim that  the unusual
features belong to the ABL rule itself, and not to the counterfactual
usage of it, because, after all, only one of the three  statements was
tested and this statement was about
actually performed experiment. I can argue against 
this claim, but instead let me discuss another aspect of this
situation.

Let us limit ourselves to the cases in which the participant claims
that the outcome of the spin component measurement, irrespectively of
the direction of measurement is $+{\hbar \over 2}$. The probability of
such a case is $1\over
4$. In this situation, in my language, there are three ``elements of
reality'': $s_x = {\hbar \over 2} $, $s_y = {\hbar \over 2} $, $s_z =
{\hbar \over 2} $. These elements of reality yield new information
about the system that the ``weak values'' (Aharonov and Vaidman, 1990) of the spin components are
$(s_x)_w = (s_y)_w =(s_z)_w  = {\hbar \over 2} $,  the proof is given
in Aharonov and Vaidman (1991). Weak values are
measured using standard measuring devices, but with weakened
coupling. This allows measuring several variables together, however, for the price
of the accuracy of the measurement. Since, for any variables, $(A+B)_w
= A_w +B_w$, the weak value of the spin
component $s_{\xi}\equiv {1\over \sqrt 3}(s_x + s_y + s_z)$ is
$(s_{\xi})_w ={{{\sqrt 3}\hbar} \over 2}$ (while the eigenvalues of $s_{\xi}$ are
$\pm {\hbar \over 2}$).
 The  center of the
distribution of the particle position in the Stern-Gerlach experiment
measuring $s_{\xi}$ using weak coupling will be outside the range of the eigenvalues!
 This is the new and nontrivial
information which we learn about the system characterized by the
three elements reality (the ABL counterfactuals) stated
above.

\vskip 1.8cm
 \noindent
{\bf 6. The three-boxes example.}
\vskip .6cm

Another example mentioned by Kastner, the particle in three boxes
pre-selected in a superposition of being in all three boxes ${1\over
  {\sqrt 3}}(|A\rangle + |B\rangle + |C\rangle)$ and post-selected in another
superposition in all three boxes ${1\over {\sqrt 3}}(|A\rangle + |B\rangle -
|C\rangle)$. The surprising feature of this particle is that we are certain to
find it inside box $A$ if it searched there and also inside box $B$ if
it searched there instead.

It is not  trivial as Kastner's raffle example, since neither
pre-selection nor post-selection alone specify the truth of the
counterfactual. Moreover, it is more subtle than a trivial example of
this kind with a particle pre-selected in $A$ and $B$ and
post-selected in $B$ and $C$  which is to be found with  certainty inside box
$B$ if it is searched there. 

In our example, the particle acts as if it is simultaneously in two
boxes for any single test of its location. As for the
spin-$1\over 2$ particle example, even more interesting features of
the quantum system 
which yield the elements of reality: ``the particle is in $A$'', and
``the particle is in $B$'' find their manifestation  in the results of weak
measurements (Vaidman, 1999c).  The particle acts as if it is
simultaneously in two boxes for multiple weak measurements  of its
location.  Recently, Aharonov and I (2002) noticed yet another
surprising feature of the system with these counterfactuals.
The particle acts as if it is simultaneously in two boxes also for a
``superposition'' of strong tests of its location.  An external single
quantum particle in a superposition of moving toward boxes $A$ and $B$
which interacts {\em strongly} with the particle in the box will
scatter from our particle in the way as if there were two particles,
one in each box.  (Note that Kastner (2002) has not found this example
surprising either.)

\vskip 1.8cm
 \noindent
{\bf 7. Time asymmetry of counterfactuals.}
\vskip .6cm

Beyond the controversy about the level of triviality (or
non-triviality) of particular examples, the ABL counterfactuals
can play an important role in a more general controversy about time
asymmetry  of all counterfactuals. There is a trend to view
counterfactuals invariably time asymmetric. For example, Kutach (2002)
worries that his Entropy Theory of Counterfactuals ``fails as an
explanation of counterfactual asymmetry''. 

 According to the orthodox view,
determining the truth  values of counterfactuals is finding the most
similar worlds where the antecedent holds, but usually there are no
well defined criteria for similarity of worlds.
 The situation is more 
clear when we consider counterfactuals related to
behavior of a physical system under changed conditions at time $t$: In
order to evaluate the counterfactual we have to consider possible
worlds which are similar to the actual world except for the antecedent of
the counterfactual. In asymmetric counterfactual, in which the similarity
of the worlds is considered only in the past, the worlds of the physical
system  can be considered {\em identical}. The change at time $t$
caused by an external intervention. (In general-type counterfactuals
in which there is no division into the system and an external agent,
we have to deal with the question of the origin of the change.)
However, in time-symmetric counterfactuals in which  
the system is considered both before and after the intervention at
time $t$, it seems that we cannot consider identical worlds. In
particular, it seems to be the case for a quantum measurement at time
$t$, because  measurements change the state of quantum systems.\footnote{\baselineskip 20pt  This
was the point of the paragraph quoted by Kastner. It is different from
the difficulty of fixing the outcomes of the measurements at two
times, which Kastner discussed. Even when these outcomes are fixed, the
state of the quantum system depends on the type of the measurement at
time $t$.}

My definition of counterfactuals (ii) resolves this difficulty (Vaidman 1999a, 1999b). The
counterfactual and actual worlds are {\it identical} except for what
is happening at $t$, the time at which the actual and the
counterfactual worlds differ by definition.  The solution came from
defining a world of a quantum system by the list of the results of
measurements performed on that system. It also fits well in the
framework of the many worlds interpretation of quantum mechanics where
I defined the concept of a world in a similar manner (Vaidman, 2002).

As far as I know, this is the first example of a nontrivial
time-symmetric counterfactual and its existence might change the trend
of considering counterfactuals as necessarily time-asymmetric.

  This research was supported in part by grant 62/01 of the
Israel Science Foundation.

\vskip 1.8cm
 \noindent
{\bf References}
\vskip .6cm

%\begin{thebibliography}{9}
 
\vskip .13cm \noindent 
Aharonov, Y.,  Bergmann,  P.G., and  Lebowitz, J.L. (1964),
``Time Symmetry in the Quantum Process of Measurement'',
 {\em Physical Review}  B 134: 1410-1416. 
  
\vskip .13cm \noindent 
Aharonov, Y. and Vaidman, L. (1990),
 ``Properties of a Quantum System
During the Time Interval Between Two Measurements'',
{\em Physical Review}   A 41: 11-20.

\vskip .13cm \noindent 
Aharonov, Y. and Vaidman, L. (1991),
``Complete Description of a Quantum System at a Given Time'',
{\em Journal of  Physics}   A 24: 2315-2328.

\vskip .13cm \noindent 
Aharonov, Y. and Vaidman, L. (2002),
``How One Shutter Can Close $N$ Slits'', e-print, quant-ph/0206074.

\vskip .13cm \noindent  Einstein, A.,  Podolsky, B., and  Rosen,
N. (1935),
``Can Quantum-Mechanical Description of Physical Reality Be Considered
Complete?'', {\it Physical  Review} 47: 777-780.

\vskip .13cm \noindent 
Kastner, R. E. (1999a),
``Time-Symmetrized Quantum Theory, Counterfactuals,
and `Advanced Action'~'',
{\em Studies in History and Philosophy of Modern Physics}  30: 237-259.

\vskip .13cm \noindent 
Kastner, R. E. (1999b),
``The Three-Box Paradox and Other Reasons to Reject the Counterfactual
Usage of the ABL Rule'', {\it Foundations of Physics} 29: 851-863.

\vskip .13cm \noindent 
Kastner, R. E. (2002),
``Shutters, Boxes, But No Paradoxes'',
 e-print, quant-ph/0207070.

\vskip .13cm \noindent 
Kastner, R. E. (2003), ``The Nature of the Controversy Over
Time-Symmetric Quantum Counterfactuals'', {\it Philosophy of Science}, March 2003,
e-print, PITT-PHIL-SCI00000868.

\vskip .13cm \noindent 
Kutach, D. N. (2002),
``The Entropy Theory of Counterfactuals'', {\it Philosophy of Science},
 69:  82-104. 

\vskip .13cm \noindent 
Sharp, W.D. and Shanks, N. (1993),
``The Rise and Fall of Time-Symmetrized Quantum Mechanics'',
{\em Philosophy of Science} 60: 488-499.

\vskip .13cm \noindent
Schulz, O., Steinhbl, R., Englert, B.G., Kurtsiefer, G., and
Weinfurter, H. (2002),
``The Mean King's Problem: Experimental Realization'', e-print,
quant-ph/0209127.

\vskip .13 cm \noindent
 Vaidman L. (1993), ``Lorentz-Invariant ``Elements of Reality" and the Joint
Measurability of Commuting Observables'',
 Physical  Review   Letters {\bf 70}, 3369-3372.
 
\break

\vskip .13cm \noindent 
 Vaidman, L. (1999a),''Defending Time-Symmetrized  Quantum  Counterfactuals'',
{\em Studies in History and Philosophy of Modern Physics}  30:
237-259, e-print version, quant-ph/9811092. 

\vskip .13cm \noindent 
 Vaidman, L. (1999b), ``Time-Symmetrized  Counterfactuals in  Quantum
 Theory'', {\it Foundations of Physics}  29: 755-765.

\vskip .13cm \noindent 
 Vaidman, L. (1999c), ``The Meaning of Elements of Reality and Quantum
 Counterfactuals -- Reply to Kastner'', {\it Foundations of Physics}  29: 865-876.

\vskip .13cm \noindent 
 Vaidman, L. (2002),
``The Many-Worlds Interpretation of Quantum Mechanics'',
{\it The Stanford Encyclopedia of
  Philosophy} (Summer 2002 Edition), Edward N. Zalta
(ed.), URL = http://plato.stanford.edu/entries/qm-manyworlds/

\vskip .13cm \noindent 
 Vaidman, L., Aharonov, Y., and Albert, D. (1987),
``How to Ascertain the Values of $\sigma_x, \sigma_y,$ and $\sigma_z$ of
a Spin-${1\over 2}$ Particle'',
{\em  Physical Review Letters} 58: 1385-1387.

\end{document}